      [1999/12/01 v1.4c Il Nuovo Cimento]
\documentclass{cimento}

\usepackage{amsmath}
\usepackage{graphicx}  
\title{GRB observations with H.E.S.S.}
\author{P.~H.~Tam\from{ins:x}\ETC,
S.~J.~Wagner\from{ins:x},
G.~P\"{u}hlhofer\from{ins:x}\\
        \atque
the H.E.S.S. Collaboration} \instlist{\inst{ins:x}Landessternwarte
Heidelberg - K\"{o}nigstuhl,
    D-69117, Heidelberg, Germany}

\PACSes{\PACSit{95.55.Ka}{X- and gamma-ray telescopes and instrumentation} \PACSit{98.70.Rz}{$\gamma$-ray sources; $\gamma$-ray bursts}}
\begin{document}

\maketitle

\begin{abstract}
H.E.S.S. (High Energy Stereoscopic System), which is designed to
detect TeV gamma-rays, is a system of four Imaging Atmospheric
Cherenkov Telescopes situated in Namibia. The system has been shown
to be very successful in detecting and observing galactic and
extra-galactic TeV sources. In order to explore the highest energy
end of GRB spectra, a GRB observing program has been established in
the H.E.S.S. collaboration. Here we introduce our GRB observing
program and report on its current status.

\end{abstract}

\section{Very-high-energy emission from GRBs}

The highest energy radiation from GRBs ever detected unambiguously
by any instrument was a $\sim18$~GeV photon coming from
GRB~940217~\cite{ref:hurley94}. In the context of standard models of
GRBs, photons with energies up to $\sim10$~TeV are expected. There
could be an energy flux radiated in this largely-unexplored energy
regime comparable to that radiated in keV-MeV or X-ray-to-radio
energies. With the high sensitivity level of H.E.S.S. (for a source
with integral flux $\sim1.4\times10^{-11}$~ph~cm$^{-2}$s$^{-1}$
above 1~TeV and spectral index 2.6, only a 2-hour observation with
H.E.S.S. is needed for a 5-$\sigma$ detection ($>$10 events)), we
are capable to detect any signal comparable to that predicted
in~\cite{ref:zhang01,ref:wang05}.

\section{H.E.S.S. GRB observing program}

A review of the system and observational highlights of H.E.S.S. can
be found in~\cite{ref:hofmann05}.

Since early 2005, the Multi-wavelength Working Group in the H.E.S.S.
collaboration has been responsible for possible GRB observations.
Every month, a member (called the contact person) from the GRB
sub-group is scheduled to be the coordinator between the team and
the shift crew at the H.E.S.S. site for any possible GRB
observations. A 24-hour-per-day monitoring is guaranteed.

We currently follow on-board GRB triggers distributed by {\it
Swift}, as well as triggers from HETE~II and INTEGRAL confirmed by
ground-based analysis. Upon the reception of a new GRB notice (with
good indications of being a true GRB) from the GRB Coordinates
Network (GCN), we observe the reported burst position as soon as
possible. The observation is limited to Z.A. $\leq$ 45 deg (to
achieve a reasonably low energy threshold) and H.E.S.S. dark time
which requires that the moon is down. The H.E.S.S. dark-time
fraction is about 0.2. When a GRB observation is being performed,
the GRB team members and the GRB contact person of that period are
responsible for the ongoing observation and keep the shift crew
updated of any new information about the GRB.

\section{Current status of the program}
We have been observing GRBs since early 2003. At the beginning of
2005, a GRB coordination team was set up and since then our GRB
observation program (see above) has been fully established. By May
2006, 14 GRB positions have been observed with H.E.S.S. (as shown in
table~\ref{tab:GRBtable}). The results will be published
elsewhere~\cite{ref:ahar07}.

\begin{table}
  \caption{14 GRBs observed with H.E.S.S. from March 2003 to May 2006}
  \label{tab:GRBtable}
  \begin{narrowtabular}{2cm}{lcccccc}
    \hline
      GRB     & Observation started after GRB onset & redshift       \\
    \hline
      060526  & !4.7 h              & 3.21! \cite{ref:berger06}          \\
      060505  & 19.4 h              & 0.089 \cite{ref:ofek06}         \\
      060403  & 13.6 h              & --             \\
      050922C & 52 min              & 2.199 \cite{ref:delia05}         \\
      050801  & 16 min              & 1.45! \cite{ref:roming06}          \\
      050726  & 10.8 h              & --             \\
      050607  & 14.8 h              & --             \\
      050509C & 21?{.}! h           & --             \\
      050209  & 20.2 h              & --             \\
      041211  & !9.5 h              & --             \\
      041006  & 10.4 h              & 0.716 \cite{ref:price04}        \\
      040425  & 26?{.}! h           & --             \\
      030821  & 18?{.}! h           & --             \\
      030329  & 11.5 d              & 0.169 \cite{ref:stanek03}        \\
      \hline
  \end{narrowtabular}
\end{table}

\acknowledgments

\end{document}